\documentstyle[12pt,epsfig]{article} 

\begin{document}  
\vspace*{-2cm}  
\renewcommand{\thefootnote}{\fnsymbol{footnote}}  
\begin{flushright}  
hep-ph/0102097\\
PSI-PR-01-03\\  
Feb. 2001\\  
\end{flushright}  
\vskip 45pt  
\begin{center}  
{\Large \bf Electroweak renormalization group corrections in high 
energy processes}\\
\vspace{1.2cm} 
{\bf  
Michael Melles\footnote{Michael.Melles@psi.ch}   
}\\  

\begin{center}
Paul Scherrer Institute (PSI), CH-5232 Villigen, Switzerland. 
\end{center}

\vspace{20pt}  
\begin{abstract}
At energies ($\sqrt{s}$) much higher than the electroweak gauge boson masses ($M$) large logarithmic
corrections of the scale ratio $\sqrt{s}/M$ occur. While the electroweak Sudakov type double (DL) and
universal single (SL) logarithms have recently been resummed, at higher orders the electroweak 
renormalization group (RG)
corrections are folded with the DL Sudakov contributions and must be included for a consistent
subleading treatment to all orders. In this paper we 
derive first all relevant formulae for massless as well as massive
gauge theories including all such terms up to order
${\cal O} \left( \alpha^n \beta_0 \log^{2n-1} \frac{s}{M^2} \right)$ by integrating over
the corresponding running couplings. The results for broken gauge theories
in the high energy regime are then given in the framework of the infrared evolution equation (IREE)
method.
The analogous QED-corrections below the weak scale
$M$ are included by appropriately matching the low energy solution to the 
renormalization group improved high energy results.
The corrections are valid for arbitrary external lines and largest in the scalar Goldstone and
Higgs boson sector as well as for transverse gauge bosons. At TeV energies, these SL-RG terms
change scattering cross sections
in the percentile regime at two loops and are thus non-negligible for precision objectives at
future linear colliders.
\end{abstract}
\end{center}  
\vskip12pt

\setcounter{footnote}{0}  
\renewcommand{\thefootnote}{\arabic{footnote}}  
  
\vfill  
\clearpage  
\setcounter{page}{1}  
\pagestyle{plain} 
 
\section{Introduction} 

With the advent of colliders in the TeV regime there has been a renewed interest in the
high energy predictions of the Standard Model (SM). At hadronic colliders the 
experimental and/or theoretical accuracy is usually in the few percent regime, and thus
the effect of one loop electroweak corrections of the order of ${\cal O} \left( 10-20 \% \right)$
at TeV energies is indeed relevant for many processes. The reason for these large
corrections to physical cross sections is primarily that they depend on the infrared cutoff,
i.e. the gauge boson masses ($M$), leading to large DL and SL corrections of the scale ratio
$\sqrt{s}/M$. Only soft photon effects need to be considered in a semi-inclusive way but
even fully inclusive cross sections are expected to depend on $\log \frac{s}{M^2}$ terms
\cite{ccc} due to the fact that the initial states carry a non-Abelian group charge
(the weak isospin) and thus violate the Bloch-Nordsieck theorem.

At this point all experimental constraints indicate that a light Higgs particle 
below the $W^\pm$ threshold is responsible
for the breaking of the electroweak symmetry. If this scenario is realized in nature 
new physics is generally expected around the TeV scale in order to avoid the hierarchy
problem. The high precision measurements of SLC/LEP have limited the room for extensions
of the SM considerably and in general, they cannot deviate from the SM to a large extent
without evoking so-called conspiracy effects. It would therefore be very desirable to 
have a leptonic collider at hand in the future in order to answer questions posed by
discoveries made at the LHC and possibly the Tevatron. In particular, if only a light Higgs
is discovered, say at 115 GeV, then it is mandatory to investigate all its properties
in detail to experimentally establish the Higgs mechanism including a possible reconstruction
of the potential and of course of the Yukawa couplings. In addition one would have to
look for additional heavy Higgs-bosons which could easily escape detection at the hadronic
machines, but have a better chance for instance at the $\gamma \gamma$-option at TESLA
\cite{kmsz,gko,mks}.
If any supersymmetric particle would be found in addition, it is necessary to
clarify and/or test the relations between couplings and properties of all new particles in
as much detail as possible in a complementary way to what would already be known by that
time.
The overall importance of leptonic colliders would thus be to clarify the physics responsible
for the electroweak symmetry breaking which in turn means it must be a high precision
machine. 

On the theory side this means that effects at the 1 \% level should be under control in both
the SM as well as all extensions that are viable at that point.
The focus of the present work is the former. In particular the abovementioned large DL
and SL corrections in the SM can, at two loops, be of the order of a few \%.
The largest contribution in the high energy limit, the DL corrections, were treated 
comprehensively to all orders in Ref. \cite{flmm}.
The method employed in Ref. \cite{flmm} is based on a non-Abelian generalization of a
bremsstrahlung theorem due to Gribov \cite{grib}. The essential point here is that corrections
factorize with respect to the perpendicular Sudakov component $|{\mbox{\boldmath $k$}_{\perp }}|$ 
of the exchanged gauge boson.
With a cutoff imposed on the allowed values of $|{\mbox{\boldmath $k$}_{\perp }}| \geq \mu \geq
M$ all gauge
bosons in the unbroken (high energy) regime of the electroweak theory factorize according
to the underlying $SU_L(2) \times U_Y(1)$ symmetry in analogy to QCD. The effect of soft photon
emission can then be included in the framework of the IREE method \cite{kl} with appropriate
matching conditions. This approach was extended to the subleading level and to longitudinal
degrees of freedom via the equivalence theorem 
in Refs. \cite{m1,m3} by employing the virtual contributions to the
respective splitting functions. In Ref. \cite{m3} it was furthermore shown that also 
top-Yukawa enhanced subleading corrections can be included in this formalism to SL accuracy
to all orders. These terms are typical for broken gauge theories as are longitudinal degrees
of freedom in general.

At one loop, the approach was tested with calculations in the physical SM fields of Refs.
\cite{dp,bddms}.
Also at the two loop level to DL accuracy, the approach was verified by explicit 
calculations in Refs. \cite{m2,bw,hkk}. To subleading accuracy it agrees with the result
of Ref. \cite{kps} for $e^+e^- \rightarrow f \overline{f}$
to all orders (up to Yukawa terms), where results for the QCD form
factor were generalized to the electroweak theory in a similar spirit as detailed above.
In addition, non-universal angular terms were calculated at the one loop level and
it was proposed to resum these terms by multiplying these corrections with the DL form factor.

The still outstanding corrections of the universal, i.e. process independent, type
are the focus of this work. They are given by the folding of DL-corrections with RG loops
at higher orders, starting at the two loop level. These contributions are of order
${\cal O} \left( \alpha^n \beta_0 \log^{2n-1} \frac{s}{M^2} \right)$ and as such
need to be included in a genuinely SL analysis. We will denote
them as SL-RG in the following. Conventional RG corrections, however,
are subsubleading at the two loop level.

The paper is organized as follows.
In section \ref{sec:qcd} we review the case for unbroken gauge theories and focus on QCD in
particular. We derive analytical formulas for both virtual and real corrections to external
quark and gluon lines depending on the experimental requirements. Section \ref{sec:sm}
then applies the results according to the above considerations to the SM after briefly
summarizing the results for the Sudakov corrections. We discuss the size of the results in section
\ref{sec:dis} and make concluding remarks in section \ref{sec:con}.

\section{Higher order renormalization group corrections in QCD} \label{sec:qcd} 

\begin{figure}
\centering
\epsfig{file=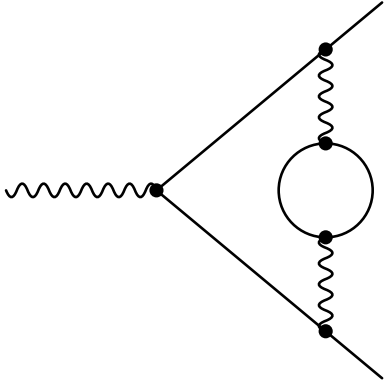,width=5cm}
\caption{A QED diagram at the two loop level yielding a SL-RG correction. The explicit result
obtained in Ref. \cite{bur} for the case
of equal masses relative to the Born amplitude
was $-\frac{1}{36} \frac{e^4}{16 \pi^4} \log^3 \frac{s}{m^2}
=\frac{1}{12} \beta_0^{\rm QED} \frac{e^4}{16 \pi^4} \log^3 \frac{s}{m^2}$. This result
is reproduced exactly by including a running coupling into the one loop vertex
correction diagram. The argument of the coupling must depend on the 
component of the loop momentum (going into the fermion loop) which is perpendicular to the
external fermion momenta. In QCD, although more diagrams contribute, the net effect is just
to replace $\beta_0^{\rm QED} \longrightarrow \beta_0^{\rm QCD}$ in the above expression.}
\label{fig:SLRG}
\end{figure}

In this section we review the case of unbroken gauge theories like QCD. Explicit comparisons with
higher order calculations revealed that the relevant RG scale in the respective diagrams is
indeed the perpendicular Sudakov component \cite{b,ddt,cl}. 
We give correction factors for each external line below. The universal nature of the higher
order SL-RG corrections can be seen as follows. Consider the gauge invariant fermionic part
($\sim n_f$) as indicative of the full $\beta^{\rm QCD}_0$ term (replacing $n_f = \frac{3}{T_F} \left(
\frac{11}{12} C_A - \beta^{\rm QCD}_0 \right) $). In order to lead to subleading, i.e. ${\cal O} \left(
\alpha^n_s \log^{2n-1} \frac{s}{\mu^2} \right)$, this loop correction must be folded
with the exchange of a gauge boson between two external lines (producing a DL type contribution)
like the one depicted in Fig. \ref{fig:SLRG}. 
Using the conservation of the total non-Abelian group charge, i.e.
\begin{equation}
\sum_{j=1}^n T^a(j) {\cal M}(p_1,...,p_j,...,p_n;{\mbox{\boldmath
$k$}_{\perp }^{2}})=0
\end{equation}
the double sum over all external insertions
$j$ and $l$ is reduced to a single sum over all $n$ external legs.
Thus these types of
corrections can be identified with external lines at higher orders. The same conclusion
is reproduced by the explicit pole structure of $\overline{\rm MS}$ renormalized scattering
amplitudes at the two loop level in QCD \cite{cat}.
In addition, from the expression in Ref. \cite{cat} it can be seen that the SL-RG corrections are
independent of the spin, i.e. for both quarks and gluons the same running coupling argument is
to be used. This is a consequence of the fact that these corrections appear only in loops
which can yield DL corrections on the lower order level and as such, the available DL phase
space is identical up to group theory factors.
We begin with the virtual case.

\subsection{Virtual corrections}

The case of virtual SL-RG corrections for both massless and massive partons has been discussed in
Ref. \cite{ms3} with a different Sudakov parametrization. Below we show the identity of both
approaches.
The form of the corrections is given in terms of the probabilities $W_{i_V} \left( s,\mu^2
\right)$. To logarithmic accuracy, they correspond to the probability to emit a soft and/or
collinear virtual parton from particle $i$ at high energies subject to an infrared
cutoff $\mu$. At the amplitude level all expressions below are
universal for each external line and exponentiate according to 
\begin{equation}
{\cal M} (p_1,...,p_n,g_s, \mu) = {\cal M}_{\rm Born} (p_1,...,p_n,g_s) \exp \left(- \frac{1}{2}
\sum_{i=1}^n W_{i_V} \left( s,\mu^2 \right) \right)
\end{equation}
where $n$ denotes the number of external lines. We begin with the massless case.

\subsubsection{Massless QCD}

In the following we denote the running QCD-coupling by
\begin{equation}
\alpha_s ({\mbox{\boldmath $k$}_{\perp }^{2}})
 = \frac{\alpha_s(\mu^2)}{1+ \frac{\alpha_s(\mu^2)
 }{\pi} \beta^{\rm QCD}_0 \log \frac{{\mbox{\boldmath $k$}_{\perp }^{2}}}{\mu^2}}
 \equiv \frac{\alpha_s(\mu^2)}{1+c \; \log \frac{{\mbox{\boldmath $k$}_{\perp }^{2}}}{\mu^2}} 
 \label{eq:rc}
 \end{equation}
 where $\beta^{\rm QCD}_0=\frac{11}{12} C_A -\frac{1}{3} T_F n_f$
 and for QCD we have
 $C_A=3$, $C_F=\frac{4}{3}$ and $T_F=\frac{1}{2}$ as usual.
 Up to two loops the massless $\beta$-function is independent of the chosen
 renormalization scheme and is gauge invariant in minimally subtracted schemes
 to all orders \cite{c}. These features will also hold for the derived
 renormalization group correction factors below in the high energy regime.
 The scale $\mu$ denotes the infrared cutoff on the exchanged ${\mbox{\boldmath $k$}_{\perp }}$
 between the external momenta $p_j,p_l$, where the Sudakov decomposition is given by
 $k=vp_l+up_j + k_\perp$, such that $p_jk_\perp =p_lk_\perp =0$. The cutoff $\mu$ serves
 as a a lower limit on the exchanged Euclidean component $\mbox{\boldmath $k$}^2_\perp=
 -k^2_\perp > 0$ which
 can be defined
 in an invariant way as:
\begin{equation}
 \mu^2 \leq \mbox{\boldmath $k$}^2_\perp \equiv \min (2 (kp_l)(kp_j)/(p_lp_j))
 \end{equation}
 for all $j \neq l$. In order to avoid the Landau pole we must choose $\mu > \Lambda_{\rm QCD}$.
 Thus, the expressions given in this section correspond for quarks to the case where $m \ll \mu$.
 For arbitrary external lines we then have
\begin{equation}
{\widetilde W}^{\rm DL}_{i_V} \left(s,\mu^2 \right) =  \frac{\alpha_s C_i}{2\pi} 
\int^s_{\mu^2} \frac{d {\mbox{\boldmath $k$}_{\perp }^{2}}}{
{\mbox{\boldmath $k$}_{\perp }^{2}}} \int^1_{{\mbox{\boldmath $k$}_{\perp }^{2}}/s}
\frac{d v}{v} =  \frac{\alpha_s C_i}{4\pi} \log^2 \frac{s}{\mu^2} 
 \end{equation}
The RG correction is then described by including the effect of the running coupling from the
scale $\mu^2$ to $s$ according to \cite{b,ddt,cl} (see also discussions in
Refs. \cite{ms3,mdur}):
\begin{eqnarray}
{\widetilde W}^{\rm RG}_{i_V}\left(s,\mu^2 \right) &=&  \frac{C_i}{2\pi} 
\int^s_{\mu^2} \frac{d {\mbox{\boldmath $k$}_{\perp }^{2}}}{
{\mbox{\boldmath $k$}_{\perp }^{2}}} \int^1_{{\mbox{\boldmath $k$}_{\perp }^{2}}/s}
\frac{d v}{v}  
 \frac{\alpha_s(\mu^2)}{1+c \; \log \frac{{\mbox{\boldmath $k$}_{\perp }^{2}}}{\mu^2}} \nonumber \\
&=& \frac{\alpha_s(\mu^2) C_i}{2 \pi } \left\{ \frac{1}{c} \log \frac{s}{\mu^2}
\left( \log \frac{\alpha_s(\mu^2)}{\alpha_s
(s)} - 1 \right) + \frac{1}{c^2}
\log \frac{\alpha_s(\mu^2)}{\alpha_s(s)} \right\} \label{eq:vrg}
 \end{eqnarray}
where $C_i=C_A$ for gluons and $C_i=C_F$ for quarks.
For completeness we also give the
subleading terms of the external line correction which is of course also
important for phenomenological applications. The terms depend on the external line
and the complete result
to logarithmic accuracy is given by:
\begin{eqnarray}
W^{RG}_{g_V}\left(s,\mu^2 \right)
&=& \frac{\alpha_s(\mu^2) C_A}{2 \pi } \left\{ \frac{1}{c} \log \frac{s}{\mu^2}
\left( \log \frac{\alpha_s(\mu^2)}{\alpha_s
(s)} - 1 \right) + \frac{1}{c^2}
\log \frac{\alpha_s(\mu^2)}{\alpha_s(s)} \right. \nonumber \\
&& \left.
 -\frac{2}{C_A} \beta^{\rm QCD}_0 \log \frac{s}{\mu^2}
\right\} \label{eq:gvrgex} \\ 
W^{RG}_{q_V}\left(s,\mu^2 \right)
&=& \frac{\alpha_s(\mu^2) C_F}{2 \pi } \left\{ \frac{1}{c} \log \frac{s}{\mu^2}
\left( \log \frac{\alpha_s(\mu^2)}{\alpha_s
(s)} - 1 \right) + \frac{1}{c^2}
\log \frac{\alpha_s(\mu^2)}{\alpha_s(s)} \right. \nonumber \\
&& \left.
 -\frac{3}{2} \log \frac{s}{\mu^2}
\right\} \label{eq:qvrgex}
 \end{eqnarray}
It should be noted that the subleading term in Eq. (\ref{eq:gvrgex}) proportional to $\beta^{\rm
QCD}_0$
is not a conventional renormalization group corrections but rather an anomalous scaling dimension,
and enters with the opposite sign \cite{m1} compared to the conventional RG contribution.

\subsubsection{Massive QCD}

Here we give results for the case when the infrared cutoff $\mu \ll m$, where $m$ denotes
the external quark mass. We begin with the case of equal external and internal line masses:
\begin{figure}
\centering
\epsfig{file=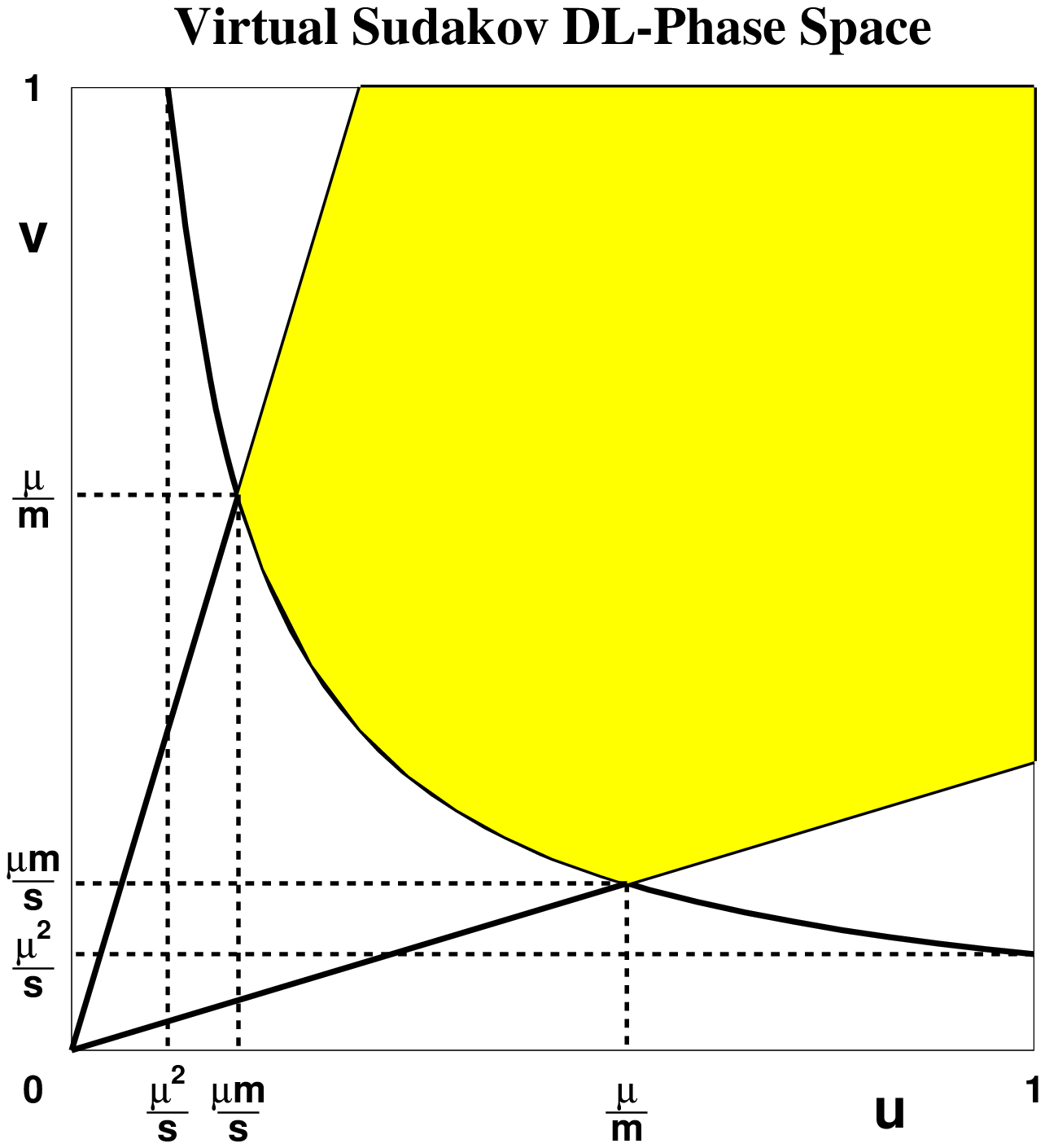,width=7.6cm}
\epsfig{file=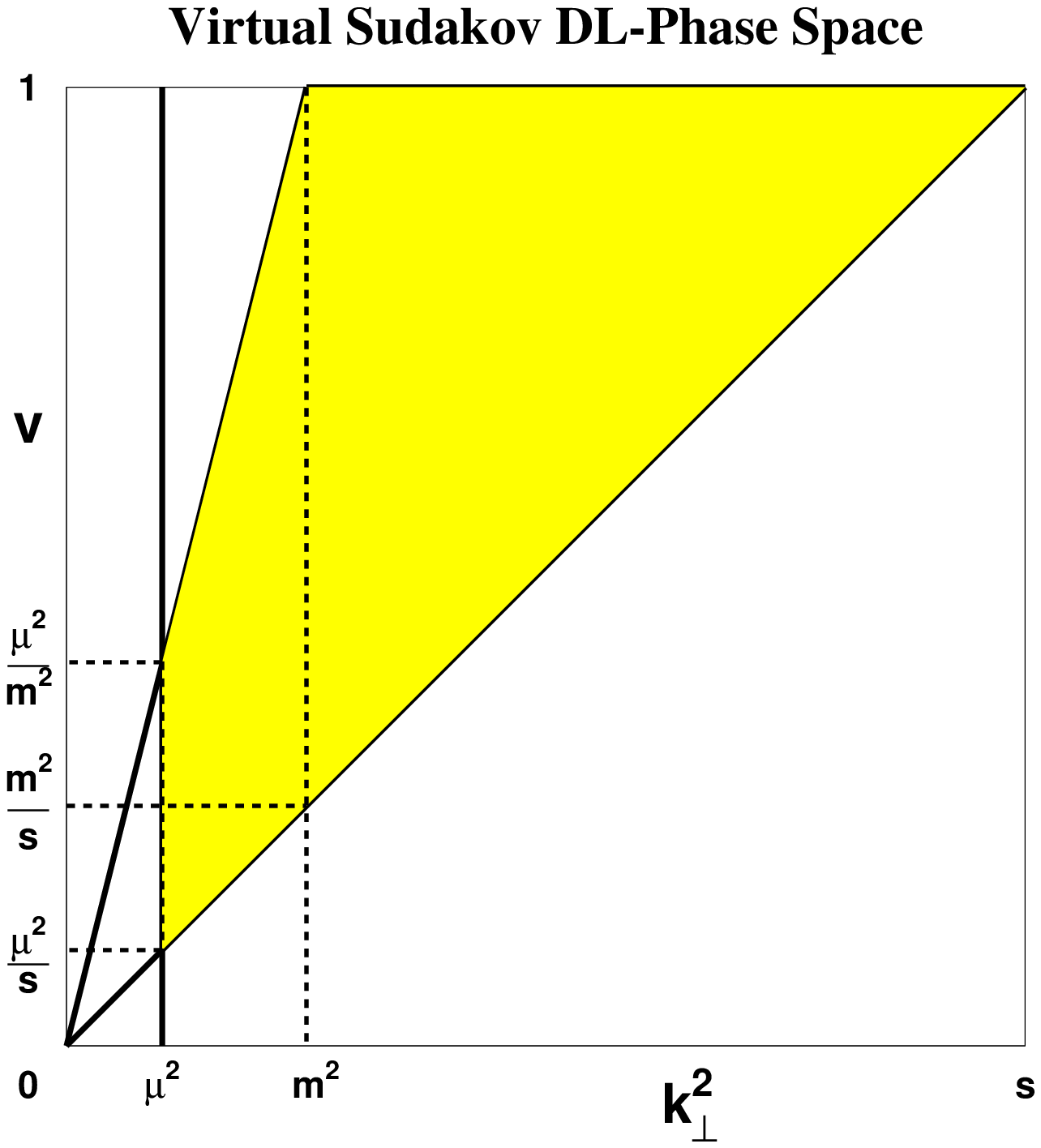,width=7.6cm}
\caption{The virtual Sudakov DL-phase space in massive QCD in the $\{ u,v \}$ and $\{
{\mbox{\boldmath $k$}_{\perp }^2}, v \}$ representation. The shaded area is the region
of integration in each case. For $\mu \geq m$ the relevant phase space
is mass independent in each case.}
\label{fig:vsud}
\end{figure}

\subsubsection*{Equal masses}

Following Ref. \cite{ms3}, we use the gluon on-shell condition
$suv={\mbox{\boldmath $k$}_{\perp }^{2}}$ to calculate the integrals.
We begin with the correction factor for each external massive quark line. Following the
diagram on the left in Fig. \ref{fig:vsud} we find:
\begin{eqnarray}
{\widetilde W}^{RG}_{q_V} \left(s,\mu^2 \right) &=&  \frac{C_F}{2 \pi} \int^1_0
\frac{d u}{u} \int^1_0 \frac{d v}
{v} \Theta ( s uv - \mu^2) \Theta ( u
- \frac{m^2}{s} v) \Theta (v- \frac{m^2}{s} u
) \nonumber \\
&& \times \frac{\alpha_s(m^2)}{1+c \; \log \frac{s uv}{m^2}}
\nonumber \\ &=&
\frac{C_F}{2 \pi} \left\{ \int^\frac{\mu}{m}_\frac{\mu^2}{s}
\frac{d u}{u} \int^1_\frac{\mu^2}{
s u} \frac{d v}{v} +
\int^1_\frac{\mu}{m} \frac{d u}{u} \int^1_
{\frac{m^2}{s}u} \frac{d v}{v} \right. \nonumber
\\ && \left.  -\int^\frac{\mu m}{s}_\frac{\mu^2}{s} \frac{d
u}{u}
\int^1_\frac{\mu^2}{su} \frac{d v}{v}
- \int^\frac{m^2}{s}_\frac{\mu m}{s} \frac{d u}{u}
\int^1_{\frac{s}{m^2}u} \frac{d v}{v} \right\}
\frac{\alpha_s(m^2)}{1+c \;\log \frac{s uv}{m^2}} \nonumber \\
&=& \frac{\alpha_s(m^2) C_F}{2 \pi } \left\{ \frac{1}{c} \log \frac{s}{m^2}
\left( \log \frac{\alpha_s(\mu^2)}{\alpha_s
(s)} - 1 \right) + \frac{1}{c^2}
\log \frac{\alpha_s(m^2)}{\alpha_s(s)} \right\}
\label{eq:SvRG}
\end{eqnarray}
The $\mu$-dependent terms cancel out of any physical cross section (as they must) when 
real soft Bremsstrahlung contributions are added 
and $c=\alpha_s(m^2) \beta^{\rm QCD}_0 / \pi$ for massive quarks.
In order to demonstrate that the
result in Eq. (\ref{eq:SvRG}) exponentiates, we calculated in Ref. \cite{ms3} the explicit two loop
renormalization group improved massive virtual Sudakov corrections,
containing a different ``running scale'' in each loop.
It is of course also possible to use the scale ${\mbox{\boldmath $k$}_{\perp }^{2}}$ directly.
In this case we have according to the right diagram in Fig. \ref{fig:vsud}:
\begin{eqnarray}
{\widetilde W}^{RG}_{q_V}  \left(s,\mu^2 \right) &=&  \frac{C_F}{2\pi} \left[ \int^{m^2}_{\mu^2} \frac{d {\mbox{\boldmath $k$}_{\perp }^{2}}}{
{\mbox{\boldmath $k$}_{\perp }^{2}}} \int^{{\mbox{\boldmath $k$}_{\perp }^{2}}/m^2}_{
{\mbox{\boldmath $k$}_{\perp }^{2}}/s} \frac{d v}{v} +
\int^s_{m^2} \frac{d {\mbox{\boldmath $k$}_{\perp }^{2}}}{
{\mbox{\boldmath $k$}_{\perp }^{2}}} \int^1_{{\mbox{\boldmath $k$}_{\perp }^{2}}/s}
\frac{d v}{v} \right] 
 \frac{\alpha_s(m^2)}{1+c \; \log \frac{{\mbox{\boldmath $k$}_{\perp }^{2}}}{m^2}} \nonumber \\
&=& \frac{\alpha_s(m^2) C_F}{2 \pi } \left\{ \frac{1}{c} \log \frac{s}{m^2}
\left( \log \frac{\alpha_s(\mu^2)}{\alpha_s
(s)} - 1 \right) + \frac{1}{c^2}
\log \frac{\alpha_s(m^2)}{\alpha_s(s)} \right\}
 \end{eqnarray}
which is the identical result as in Eq. (\ref{eq:SvRG}). 
For completeness we also give the
subleading terms of the pure one loop form factor which is again
important for phenomenological applications. The complete result
to logarithmic accuracy is thus given by:
\begin{eqnarray}
W^{RG}_{q_V} \left(s,\mu^2 \right)
&=& \frac{\alpha_s(m^2) C_F}{2 \pi } \left\{ \frac{1}{c} \log \frac{s}{m^2}
\left( \log \frac{\alpha_s(\mu^2)}{\alpha_s
(s)} - 1 \right) + \frac{1}{c^2}
\log \frac{\alpha_s(m^2)}{\alpha_s(s)} \right. \nonumber \\
&& \left.
 -\frac{3}{2} \log \frac{s}{m^2} - \log \frac{ m^2}{
 \mu^2} \right\} \label{eq:mqvrgex}
 \end{eqnarray}
For $m=\mu$ Eq. (\ref{eq:mqvrgex}) agrees with
Eq. (\ref{eq:qvrgex}) in the previous section for massless quarks.

\subsubsection*{Unequal masses}

In this section we denote the external mass as before by $m$ and the internal mass
by $m_i$ and thus, the constant $c=\alpha_s(m^2_i) \beta_0^{\rm QCD} /\pi$. 
We consider only the case at high energies taking the first two families
of quarks as massless. The
running of all light flavors is implicit in the $n_f$ term of the
$\beta_0^{\rm QCD}$ function. The result is then given by:
\begin{eqnarray}
{\widetilde W}^{RG}_{q_V}  \left(s,\mu^2 \right) &=&  \frac{C_F}{2\pi} \left[ \int^{m^2}_{\mu^2} \frac{d {\mbox{\boldmath $k$}_{\perp }^{2}}}{
{\mbox{\boldmath $k$}_{\perp }^{2}}} \int^{{\mbox{\boldmath $k$}_{\perp }^{2}}/m^2}_{
{\mbox{\boldmath $k$}_{\perp }^{2}}/s} \frac{d v}{v} +
\int^s_{m^2} \frac{d {\mbox{\boldmath $k$}_{\perp }^{2}}}{
{\mbox{\boldmath $k$}_{\perp }^{2}}} \int^1_{{\mbox{\boldmath $k$}_{\perp }^{2}}/s}
\frac{d v}{v} \right] 
 \frac{\alpha_s(m_i^2)}{1+c \; \log \frac{{\mbox{\boldmath $k$}_{\perp }^{2}}}{m_i^2}} \nonumber \\
&=& \frac{\alpha_s(m_i^2) C_F}{2 \pi } \left\{ \frac{1}{c} \log \frac{s}{m^2}
\left( \log \frac{\alpha_s(\mu^2)}{\alpha_s
(s)} - 1 \right) \right. \nonumber \\ && \left. + \frac{1}{c}  
\log \frac{\alpha_s(m^2)}{\alpha_s(s)} \left( \frac{1}{c}+  \log \frac{m^2}{m_i^2}
\right) \right\}
 \end{eqnarray}
It is evident that the effect of unequal masses is large only for a large mass splitting.
In QCD, we always assume scales larger than $\Lambda_{\rm QCD}$ and with our assumptions we
have only the
ratio of $m_t/m_b$ leading to significant corrections.

The full subleading expression is accordingly given by:
\begin{eqnarray}
W^{RG}_{q_V} \left(s,\mu^2 \right)
&=& \frac{\alpha_s(m_i^2) C_F}{2 \pi } \left\{ \frac{1}{c} \log \frac{s}{m^2}
\left( \log \frac{\alpha_s(\mu^2)}{\alpha_s
(s)} - 1 \right)  \right. \nonumber \\ && \left. + \frac{1}{c}
\log \frac{\alpha_s(m^2)}{\alpha_s(s)} \left( \frac{1}{c}+  \log \frac{m^2}{m_i^2}
\right) -\frac{3}{2} \log \frac{s}{m^2} - \log \frac{ m^2}{
 \mu^2} \right\} \nonumber \\
&=& \frac{\alpha_s(m_i^2) C_F}{2 \pi } \left\{ \frac{1}{c} \log \frac{s}{m^2}
\left( \log \frac{\alpha_s(\mu^2)}{\alpha_s
(s)} - 1 \right)  \right. \nonumber \\ && \left. + \frac{1}{c^2} \frac{\alpha_s(m_i^2)}{
\alpha_s(m^2)}
\log \frac{\alpha_s(m^2)}{\alpha_s(s)} 
 -\frac{3}{2} \log \frac{s}{m^2} - \log \frac{ m^2}{
 \mu^2} \right\} \label{eq:miqvr}
 \end{eqnarray}
For $m=m_i$ Eq. (\ref{eq:miqvr}) agrees with
Eq. (\ref{eq:mqvrgex}) in the previous section for equal mass quarks.

If we want to apply the above result for the case of QED corrections later, then there
is no Landau pole (at low energies) 
and we can have large corrections of the form $m_b/m_e$ etc.
In this case the running coupling term is given by 
\begin{equation}
e^2({\mbox{\boldmath $k$}_{\perp }^{2}})= \frac{e^2}{1- 
\frac{1}{3} \frac{e^2}{4 \pi^2} \sum_{j=1}^{n_f
} Q_j^2 N^j_C \log
\frac{{\mbox{\boldmath $k$}_{\perp }^{2}}}{m_j^2}} \label{eq:eeff}
\end{equation}
and instead of Eq. (\ref{eq:miqvr}) we have:
\begin{eqnarray}
&& W^{RG}_{f_V} \left(s,\mu^2 \right)
= \frac{e_f^2}{8 \pi^2 } \left\{ \frac{1}{c} \log \frac{s}{m^2}
\left( \log \frac{e^2(\mu^2)}{e^2
(s)} - 1 \right) + \right. \nonumber \\ && \left.  \frac{1}{c^2}
\log \frac{e^2(m^2)}{e^2(s)} \left( 1- \frac{1}{3}\frac{e^2}{4 \pi^2}
\sum_{j=1}^{n_f } Q_j^2 N^j_C \log \frac{m^2}{m_j^2}
\right) -\frac{3}{2} \log \frac{s}{m^2} - \log \frac{ m^2}{
 \mu^2} \right\} \label{eq:mifvr}
 \end{eqnarray}
and where $c=- \frac{1}{3}\frac{e^2}{4 \pi^2} \sum_{j=1}^{n_f } Q_j^2 N^j_C$.

\subsection{Real gluon emission}

We discuss the massless and massive case separately since the structure of the divergences
is different in each case. For massive quarks we discuss two types of restrictions on the
experimental requirements, one in analogy to the soft gluon approximation. 
The expressions below exponentiate on the level of the cross section, i.e. 
for observable scattering cross sections they are of the form
\begin{eqnarray}
d \sigma (p_1,...,p_n,g_s,\mu_{\rm expt}) &=& d \sigma_{\rm Born}
(p_1,...,p_n,g_s) \times \nonumber \\ &&
\exp \left\{ \sum_{i=1}^n \left[ W_{i,R} \left( s,\mu^2,\mu^2_{\rm expt} 
\right)-W_{i,V} \left( s,\mu^2 \right) \right] \right\}
\end{eqnarray}
where the sum in the exponential is independent of $\mu$ and only depends on the cutoff
$\mu_{\rm expt}$ defining the experimental cross section.
We begin with the
massless case.

\subsubsection{Emission from massless partons}

In this section we consider the emission of real gluons with a cutoff 
${\mbox{\boldmath $k$}_{\perp }} \leq \mu_{\rm expt}$, related to the
experimental requirements. For massless partons we have at the DL level:
\begin{equation}
{\widetilde W}^{DL}_{i_R} \left(s,\mu^2,\mu^2_{\rm expt} \right) =  \frac{\alpha_sC_i}{\pi} 
\int^{\mu^2_{\rm expt}}_{\mu^2} \frac{d {\mbox{\boldmath $k$}_{\perp }^{2}}}{
{\mbox{\boldmath $k$}_{\perp }^{2}}} \int^{\sqrt s}_{|{\mbox{\boldmath $k$}_{\perp }}|}
\frac{d \omega}{\omega}  
= \frac{\alpha_sC_i}{4\pi}  \left\{ \log^2 \frac{s}{\mu^2} - \log^2 \frac{s}{\mu^2_{\rm expt}}
\right\}
 \end{equation}
and thus for the RG-improved correction:
\begin{eqnarray}
&& {\widetilde W}^{RG}_{i_R}\left(s,\mu^2,\mu^2_{\rm expt} \right) =  \frac{C_i}{\pi} 
\int^{\mu^2_{\rm expt}}_{\mu^2} \frac{d {\mbox{\boldmath $k$}_{\perp }^{2}}}{
{\mbox{\boldmath $k$}_{\perp }^{2}}} \int^{\sqrt s}_{|{\mbox{\boldmath $k$}_{\perp }}|}
\frac{d \omega}{\omega}  
 \frac{\alpha_s(\mu^2)}{1+c \; \log \frac{{\mbox{\boldmath $k$}_{\perp }^{2}}}{\mu^2}} \nonumber \\
&&= \frac{C_i\alpha_s(\mu^2)}{2\pi} \!\! \left\{ \frac{1}{c} \log \frac{s}{\mu^2} \left( \log \frac{
\alpha_s (\mu^2)}{\alpha_s (\mu^2_{\rm expt})} - 1 \! \right) - \frac{1}{c} \log \frac{\mu^2_{\rm expt}}{
s}
+ \frac{1}{c^2} \log \frac{\alpha_s (\mu^2)}{\alpha_s (\mu^2_{\rm expt})} \right\}
 \end{eqnarray}
This expression depends on $\mu$ as it must in order to cancel the infrared divergent virtual 
corrections. In fact the sum of real
plus virtual corrections on the level of the cross section is given by
\begin{eqnarray}
&& W^{RG}_{i_R} \left(s,\mu^2,\mu^2_{\rm expt} \right)-W^{RG}_{i_V}\left(s,\mu^2 \right) = 
\nonumber \\ &&
\frac{C_i}{2 \beta^{\rm QCD}_0} \!\! \left\{ \log \frac{s}{\mu^2} \log \frac{
\alpha_s (s)}{\alpha_s (\mu^2_{\rm expt})}  -  \log \frac{\mu^2_{\rm expt
}}{s}
+ \frac{1}{c} \log \frac{\alpha_s (s)}{\alpha_s (\mu^2_{\rm expt})} \right\} \nonumber \\
&=& \frac{C_i}{2 \beta^{\rm QCD}} \left( \frac{\pi}{\alpha(s) \beta^{\rm QCD}_0} 
\log \frac{ \alpha_s (s)}{\alpha_s 
(\mu^2_{\rm expt})} -  \log \frac{\mu^2_{\rm expt}}{s} \right)
 \end{eqnarray}
and thus independent of $\mu$.
The full expressions to subleading accuracy are thus:
\begin{eqnarray}
W^{RG}_{g_R}\left(s,\mu^2,\mu^2_{\rm expt} \right)
&=& \frac{C_A\alpha_s(\mu^2)}{2\pi} \!\! \left\{ \frac{1}{c} \log \frac{s}{\mu^2} \left( \log \frac{
\alpha_s (\mu^2)}{\alpha_s (\mu^2_{\rm expt})} - 1 \! \right) - \frac{1}{c} \log 
\frac{\mu^2_{\rm expt}}{s} \right. \nonumber \\ && \left.
+ \frac{1}{c^2} \log \frac{\alpha_s (\mu^2)}{\alpha_s (\mu^2_{\rm expt})} 
 -\frac{2}{C_A} \beta^{\rm QCD}_0 \log \frac{s}{\mu^2}
\right\} \label{eq:grrgex} \\ 
W^{RG}_{q_R} \left(s,\mu^2,\mu^2_{\rm expt} \right)
&=& \frac{C_F\alpha_s(\mu^2)}{2\pi} \!\! \left\{ \frac{1}{c} \log \frac{s}{\mu^2} \left( \log \frac{
\alpha_s (\mu^2)}{\alpha_s (\mu^2_{\rm expt})} - 1 \! \right) - \frac{1}{c} 
\log \frac{\mu^2_{\rm expt}}{s} \right. \nonumber \\ && \left.
+ \frac{1}{c^2} \log \frac{\alpha_s (\mu^2)}{\alpha_s (\mu^2_{\rm expt})} 
 -\frac{3}{2} \log \frac{s}{\mu^2}
\right\} \label{eq:qrrgex}
 \end{eqnarray}
All divergent ($\mu$-dependent) terms cancel when the full virtual corrections are added.

\subsubsection{Emission from massive quarks}

In the case of a massive quark, i.e. $\mu \ll m$,
the overall infrared divergence is not as severe.
This means we can discuss different requirements which all have the correct divergent
pole structure canceling the corresponding terms from the virtual contributions.
We divide the discussion in two parts as above.

\subsubsection*{Equal masses}

The constant $c=\alpha_s(m^2) \beta^{\rm QCD}_0 / \pi$ below.
We have the following expression without a running coupling:
\begin{eqnarray}
\!\!\!\!\!\!\!\!\!\! && \!\!\!\!\! W_{q_R}\left(s,\mu^2,\mu^2_{\rm expt} \right) =  \frac{\alpha_s C_F}{\pi}
\int^{\mu^2_{\rm expt}}_{\mu^2} d {\mbox{\boldmath $k$}_{\perp }^{2}}
\int^{\sqrt s}_{|{{\mbox{\boldmath $k$}_{\perp }}}|}
\frac{d \omega}{\omega} \frac{{\mbox{\boldmath $k$}_{\perp }^{2}}}{\left(
{\mbox{\boldmath $k$}_{\perp }^{2}}+ m^2/s \; \omega^2 \right)^2} \nonumber \\
\!\!\!\!\!\!\!\!\!\! &\approx& \!\!\!\!\! \left\{ \!\!\! \begin{array}{lc} \frac{\alpha_s C_F}{2 \pi} \!\!\left( 
\! \frac{1}{2} \log^2 \frac{s}{m^2} 
+ \log \frac{s}{m^2} \log \frac{m^2}{\mu^2}
- \log \frac{m^2}{\mu^2} - \frac{1}{2} \log^2 \frac{s}{\mu_{\rm expt}^2} \right) &\!\!\!\!\! , \, m \ll \mu_{\rm expt} \\
\frac{\alpha_s C_F}{2 \pi} \!\!
\left( \! \log^2 \frac{s}{m^2} 
+ \log \frac{s}{m^2} \log \frac{m^2}{\mu^2}
+ \log \frac{\mu^2}{\mu_{\rm expt}^2} - \log \frac{s}{m^2} \log 
\frac{s}{\mu_{\rm expt}^2} \right) 
&\!\!\!\!\! , \, \mu_{\rm expt} \ll m
\end{array} \right.
 \end{eqnarray}
If we want to employ a restriction analogously to the soft gluon approximation, we find
independently of the quark mass \cite{m1,m2}:
\begin{eqnarray}
\!\!\!\!\!\!\!\!\!\!&& W_{q_R}\left(s,\mu^2,\mu^2_{\rm expt} \right) =  \frac{\alpha_s C_F}{\pi}
\int^{\mu^2_{\rm expt}}_{\mu^2} d {\mbox{\boldmath $k$}_{\perp }^{2}}
\int^{\sqrt{\mu_{\rm expt}}}_{|{{\mbox{\boldmath $k$}_{\perp }}}|}
\frac{d \omega}{\omega} \frac{{\mbox{\boldmath $k$}_{\perp }^{2}}}{\left(
{\mbox{\boldmath $k$}_{\perp }^{2}}+ m^2/s \; \omega^2 \right)^2} \nonumber \\
\!\!\!\!\!\!\!\!\!\!&& \approx  
\frac{\alpha_s C_F}{2 \pi} 
\left( \frac{1}{2} \log^2 \frac{s}{m^2} 
+ \log \frac{s}{m^2} \log \frac{m^2}{\mu^2}
- \log \frac{m^2}{\mu^2}+ \log \frac{s}{\mu_{\rm expt}^2} - \log \frac{s}{m^2} \log 
\frac{s}{\mu_{\rm expt}^2} \right) 
 \end{eqnarray}
In all cases above we have not taken into account all subleading collinear logarithms related
to real gluon emission. In order
to now proceed with the inclusion of the running coupling terms it is convenient to first
consider only the DL phase space in each case. Thus we find
\begin{eqnarray}
{\widetilde W}^{RG}_{q_R}\left(s,\mu^2,\mu^2_{\rm expt} \right) \!\!&=&\!\!  \frac{\alpha_s (m^2) C_F}{2\pi}
\left( \int^{m^2}_{\mu^2} \frac{d {\mbox{\boldmath $k$}_{\perp }^{2}}}{
{\mbox{\boldmath $k$}_{\perp }^{2}}}
\log \frac{s}{m^2} + \int_{m^2}^{\mu_{\rm expt}^2} \frac{d {\mbox{\boldmath $k$}_{\perp }^{2}}}{
{\mbox{\boldmath $k$}_{\perp }^{2}}} \log \frac{s}{{\mbox{\boldmath $k$}_{\perp }^{2}}} \right)
\frac{1}{1+c \log \frac{
{\mbox{\boldmath $k$}_{\perp }^{2}}}{m^2}} \nonumber \\
\!\!&\approx&\!\!  \frac{\alpha_s(m^2) C_F}{2 \pi} \!\left[ 
\frac{1}{c} \log \frac{s}{m^2} \left( \log \frac{ \alpha_s ( \mu^2 )}{ \alpha_s ( \mu_{\rm expt}^2 )}
-1 \right) + \frac{1}{c} \log \frac{s}{\mu_{\rm expt}^2 } \right. \nonumber \\
&& \left. \;\;\;\;\;\;\;\;\;\;\;\;\;\;\;\;\;+\frac{1}{c^2} 
\log \frac{ \alpha_s (m^2)}{ \alpha_s ( \mu_{\rm expt}^2 )} \right] \; , \; m \ll
 \mu_{\rm expt} 
 \end{eqnarray}
and
\begin{eqnarray}
{\widetilde W}^{RG}_{q_R} \left(s,\mu^2,\mu^2_{\rm expt} \right) &=&  \frac{\alpha_s (m^2) C_F}{2\pi}
\int^{\mu_{\rm expt}^2}_{\mu^2} \frac{d {\mbox{\boldmath $k$}_{\perp }^{2}}}{
{\mbox{\boldmath $k$}_{\perp }^{2}}}
\log \frac{s}{m^2} 
\frac{1}{1+c \log \frac{
{\mbox{\boldmath $k$}_{\perp }^{2}}}{m^2}} \nonumber \\
&\approx& \frac{\alpha_s (m^2) C_F}{2\pi} \frac{1}{c}   
\log \frac{s}{m^2}
\log \frac{ \alpha_s ( \mu^2 )}{ \alpha_s ( \mu_{\rm expt}^2 )}  \; , \; \mu_{\rm expt} \ll m
 \end{eqnarray}
The full subleading expressions are thus given by
\begin{eqnarray}
W^{RG}_{q_R} \left(s,\mu^2,\mu^2_{\rm expt} \right) 
&\approx& \frac{\alpha_s (m^2)C_F}{2 \pi} \!\left[ \frac{1}{c} 
\log \frac{s}{m^2} \left( \log \frac{ \alpha_s ( \mu^2 )}{ \alpha_s ( \mu_{\rm expt}^2 )}
-1 \right) + \frac{1}{c} \log \frac{s}{\mu_{\rm expt}^2 } \right. \nonumber \\ && \left.
+\frac{1}{c^2} 
\log \frac{ \alpha_s (m^2)}{ \alpha_s ( \mu_{\rm expt}^2 )}
- \log \frac{m^2}{\mu^2}  
\right] \; , \; m \ll
 \mu_{\rm expt} 
 \end{eqnarray}
and
\begin{equation}
W^{RG}_{q_R} \left(s,\mu^2,\mu^2_{\rm expt} \right) 
\approx \frac{\alpha_s (m^2) C_F}{2\pi} \left[ \frac{1}{c} 
\log \frac{s}{m^2}
\log \frac{ \alpha_s ( \mu^2 )}{ \alpha_s ( \mu_{\rm expt}^2 )} 
+  \log \frac{\mu^2}{\mu_{\rm expt}^2 } \right] \; , \; \mu_{\rm expt} \ll m
 \end{equation}
In case we also impose a cut on the integration over $\omega$ we have independently of the relation
between $m$ and $\mu_{\rm expt}$ assuming only $m^2 \ll s$:
\begin{eqnarray}
{\widetilde W}^{RG}_{q_R}\left(s,\mu^2,\mu^2_{\rm expt} \right) &=&  \frac{\alpha_s (m^2) C_F}{2\pi}
\left( \int^{\frac{m^2 \mu_{\rm expt}^2}{s}}_{\mu^2} \frac{d {\mbox{\boldmath $k$}_{\perp }^{2}}}{
{\mbox{\boldmath $k$}_{\perp }^{2}}}
\log \frac{s}{m^2} + \int_{\frac{m^2 \mu_{\rm expt}^2}{s}}^{\mu_{\rm expt}^2} 
\frac{d {\mbox{\boldmath $k$}_{\perp }^{2}}}{
{\mbox{\boldmath $k$}_{\perp }^{2}}} \log \frac{\mu_{\rm expt}^2}{{\mbox{\boldmath $k$}_{\perp }^{2}}} \right)
\nonumber \\ && \times
\frac{1}{1+c \log \frac{
{\mbox{\boldmath $k$}_{\perp }^{2}}}{m^2}} \nonumber \\
&\approx&  \frac{\alpha_s (m^2) C_F}{2 \pi}  \left[ \frac{1}{c} \log \frac{s}{m^2}
 \left( \log \frac{ \alpha_s (\mu^2)}{\alpha_s (\mu_{\rm expt}^2)} - 1  \right) \right.
 \nonumber \\ && \left. + \frac{1}{c}
\log \frac{
s}{\mu_{\rm expt}^2} \log \frac{ \alpha_s(\mu^2_{\rm expt})}{ \alpha_s (\mu^2_{\rm expt}m^2/s)} + 
 \frac{1}{c^2} \log \frac{
\alpha_s(\mu^2_{\rm expt}m^2/s)}{\alpha_s(\mu_{\rm expt}^2)} \right] \label{eq:rem}
 \end{eqnarray}
This expression agrees with the result obtained in Ref. \cite{ms3} where the gluon on-shell
condition ${\mbox{\boldmath $k$}_{\perp }^{2}}=suv$ was used and one integral over one Sudakov
parameter was done numerically. In Ref. \cite{ms3} it was also shown that the RG-improved
virtual plus soft form factor also
exponentiates by explicitly calculating the two loop RG correction with each loop containing
a running coupling of the corresponding ${\mbox{\boldmath $k$}_{\perp }^{2}}$.

The full subleading expression for the RG-improved soft gluon emission correction is thus given by
\begin{eqnarray}
\!\!\!&&  W^{RG}_{q_R} \left(s,\mu^2,\mu^2_{\rm expt} \right)
\approx  \frac{\alpha_s (m^2) C_F}{2 \pi}  \left[ \frac{1}{c} \log \frac{s}{m^2}
 \left( \log \frac{ \alpha_s (\mu^2)}{\alpha_s (\mu_{\rm expt}^2)} - 1  \right) \right. 
 \nonumber \\ \!\!\!&& \left. + \frac{1}{c}
\log \frac{
s}{\mu_{\rm expt}^2} \log \frac{ \alpha_s(\mu^2_{\rm expt})}{ \alpha_s (\mu^2_{\rm expt}m^2/s)} +
 \frac{1}{c^2} \log \frac{
\alpha_s(\mu^2_{\rm expt}m^2/s)}{\alpha_s(\mu_{\rm expt}^2)} 
- \log \frac{m^2}{\mu^2} + \log \frac{s}{\mu_{\rm expt}^2}
\right]
 \end{eqnarray}
for the equal mass case. The case of different external and internal masses is again important
for applications in QED and will be discussed next.

\subsubsection*{Unequal masses}

While the gluonic part of the $\beta$-function remains unchanged we integrate again only
from the scale of the massive fermion which is assumed to be in the perturbative regime.
For applications to QED, however, we need the full expressions below.
Here we discuss only the case analogous to the soft gluon approximation. Considering again
only the high energy scenario we have
for the case of an external mass $m$ and a fermion loop mass $m_i$:
\begin{eqnarray}
{\widetilde W}^{RG}_{q_R}\left(s,\mu^2,\mu^2_{\rm expt} \right) &=&  \frac{\alpha_s (m_i^2) C_F}{2\pi}
\left( \int^{\frac{m^2 \mu_{\rm expt}^2}{s}}_{\mu^2} \frac{d {\mbox{\boldmath $k$}_{\perp }^{2}}}{
{\mbox{\boldmath $k$}_{\perp }^{2}}}
\log \frac{s}{m^2} + \int_{\frac{m^2 \mu_{\rm expt}^2}{s}}^{\mu_{\rm expt}^2} 
\frac{d {\mbox{\boldmath $k$}_{\perp }^{2}}}{
{\mbox{\boldmath $k$}_{\perp }^{2}}} \log \frac{\mu_{\rm expt}^2}{{\mbox{\boldmath $k$}_{\perp }^{2}}} \right)
\nonumber \\ && \times
\frac{1}{1+c \log \frac{
{\mbox{\boldmath $k$}_{\perp }^{2}}}{m^2_i}} \nonumber \\
&\approx&  \frac{\alpha_s (m^2_i) C_F}{2 \pi}  \left[ \frac{1}{c} \log \frac{s}{m^2}
 \left( \log \frac{ \alpha_s (\mu^2)}{\alpha_s (\mu_{\rm expt}^2m^2/s)} - 1  \right) \right.
 \nonumber \\ && \left. + \frac{1}{c}
\log \frac{
m_i^2}{\mu_{\rm expt}^2} \log \frac{ \alpha_s(\mu^2_{\rm expt})}{ \alpha_s (\mu^2_{\rm expt}m^2/s)} + 
 \frac{1}{c^2} \log \frac{
\alpha_s(\mu^2_{\rm expt}m^2/s)}{\alpha_s(\mu_{\rm expt}^2)} \right] \label{eq:reMm}
 \end{eqnarray}
This expression agrees with the result obtained in Eq. (\ref{eq:rem}) for the case $m_i=m$.

The full subleading expression for the RG-improved soft gluon emission correction is thus given by
\begin{eqnarray}
\!\!\!&&  W^{RG}_{q_R} \left(s,\mu^2,\mu^2_{\rm expt} \right)
\approx  \frac{\alpha_s (m^2_i) C_F}{2 \pi}  \left[ \frac{1}{c} \log \frac{s}{m^2}
 \left( \log \frac{ \alpha_s (\mu^2)}{\alpha_s (\mu_{\rm expt}^2m^2/s)} - 1  \right) \right. 
 \nonumber \\ \!\!\!&& \left. + \frac{1}{c}
\log \frac{m_i^2
}{\mu_{\rm expt}^2} \log \frac{ \alpha_s(\mu^2_{\rm expt})}{ \alpha_s (\mu^2_{\rm expt}m^2/s)} +
 \frac{1}{c^2} \log \frac{
\alpha_s(\mu^2_{\rm expt}m^2/s)}{\alpha_s(\mu_{\rm expt}^2)} 
- \log \frac{m^2}{\mu^2} + \log \frac{s}{\mu_{\rm expt}^2}
\right] \nonumber \\
&& =  \frac{\alpha_s (m^2_i) C_F}{2 \pi}  \left[ \frac{1}{c} \log \frac{s}{m^2}
 \left( \log \frac{ \alpha_s (\mu^2)}{\alpha_s (\mu_{\rm expt}^2m^2/s)} - 1  \right) \right. 
 \nonumber \\ \!\!\!&& \left. 
 + \frac{1}{c^2} \frac{\alpha_s(m_i^2)}{\alpha_s(\mu^2_{\rm expt})} \log \frac{
\alpha_s(\mu^2_{\rm expt}m^2/s)}{\alpha_s(\mu_{\rm expt}^2)} 
- \log \frac{m^2}{\mu^2} + \log \frac{s}{\mu_{\rm expt}^2}
\right] \label{eq:sg}
 \end{eqnarray}
As mentioned above, this expression is more useful for applications in QED or if the mass
ratios are very large. In QED we have again the running coupling of the form given in
Eq. (\ref{eq:eeff}), and Eq. (\ref{eq:sg}) becomes
\begin{eqnarray}
\!\!\!&&  W^{RG}_{f_R} \left(s,\mu^2,\mu^2_{\rm expt} \right)
\approx  \frac{e_f^2 }{8 \pi^2}  \left[ \frac{1}{c} \log \frac{s}{m^2}
 \left( \log \frac{ e^2 (\mu^2)}{e^2 (\mu_{\rm expt}^2m^2/s)} - 1  \right) \right. 
 \nonumber \\ \!\!\!&&  + \frac{1}{c^2} \log \frac{e^2(\mu^2_{\rm expt}m^2/s)}{e^2(\mu_{\rm expt}^2)} 
\left( 1-
 \frac{1}{3} \frac{e^2}{4 \pi^2} \sum^{n_f}_{j=1}
 Q^2_j N_C^j
\log \frac{
\mu_{\rm expt}^2}{m_j^2} \right)
\nonumber \\ && \left.
- \log \frac{m^2}{\mu^2} + \log \frac{s}{\mu_{\rm expt}^2}
\right] \label{eq:sp}
 \end{eqnarray}
where again $c=-\frac{1}{3} \frac{e^2}{4 \pi^2} \sum^{n_f}_{j=1}Q^2_j N_C^j$.
This concludes the discussion of SL-RG effects in QCD. As a side remark we mention that
for scalar quarks, the same function appears as for fermions since the DL-phase space 
for both cases is identical. Only $\beta_0$ differs in each case.

\section{Electroweak RG corrections} \label{sec:sm} 

We now turn to the case of spontaneously broken gauge theories. 
As in the previous section, we are interested only in terms of SL-accuracy.
At one loop, these are the obvious RG corrections from the running couplings as discussed
in Refs. \cite{dp,m1,m3}. At higher orders, we have the same situation as in QCD that the
RG corrections are folded into loops which on a lower order lead to DL contributions.
We begin with a summary of the known higher order DL and SL corrections.

\subsection{Subleading electroweak Sudakov corrections to all orders}

In this section we are going to discuss the higher order Sudakov corrections in the
electroweak theory. While the method discussed
below is general, for definiteness we consider only the SM.
The framework we use in the following is given by the infrared evolution equation (IREE) method.
The basic physical idea behind this framework is to identify the effective high
energy theory at values of ${\mbox{\boldmath $k$}_{\perp }} \gg \mu \geq M$. The corresponding
contribution from QED below the scale $M$ is then given by appropriate matching conditions
at $\mu=M$ in order to recover the high energy theory solution. In this way all universal
Sudakov DL and SL have been resummed in Refs. \cite{flmm,m1,m3}. At one loop the results obtained
by the IREE method agree with the literature for all external lines and at two loops, the
DL-results were checked by explicit calculations with the physical SM fields \cite{m2,bw,hkk}.
Including soft bremsstrahlung with a cut
on the allowed ${\mbox{\boldmath $k$}_\perp} \leq \mu_{\rm expt} \leq M$ of the emitted
real photons, and regularizing virtual IR divergences with a cutoff
${\mbox{\boldmath $k$}_\perp} \geq \mu$, we find for the semi-inclusive cross 
section\footnote{We emphasize that for photon and Z-boson final states
the mixing effects have to be included correctly as described in Ref. \cite{m1}. In particular,
for transverse degrees of freedom the corrections don't factorize with respect to the
physical Born amplitude
but rather with respect to the amplitudes containing the fields in the broken phase.
For longitudinally polarized Z-bosons, however, there is no mixing with photons
and the corrections factorize with respect to the Born amplitude.}:
\begin{eqnarray}
&& d\sigma (p_{1}, \ldots, p_{n},g,g^\prime,\mu_{exp}) = d\sigma_{\rm Born} (p_{1},
\ldots ,p_{n},g,g^\prime )
\nonumber \\ && \times \exp \left\{ - \sum^{n_g}_{i=1} W_{g_i} (s,M^2)
- \sum^{n_f}_{i=1} W_{f_i} (s,M^2) - \sum^{n_\phi}_{i=1} W_{\phi_i} (s,M^2)
\right\} \nonumber \\
&&\times \exp \left[ - \sum_{i=1}^{n_f} \left( w_{f_i}(s,\mu^2)
- w_{f_i}(s,M^2) \right)
- \sum_{i=1}^{n_{\rm w}} \left( w_{{\rm w}_i}(s,\mu^2)
- w_{{\rm w}_i}(s,M^2) \right) \right. \nonumber \\
&& \;\;\;\;\;\;\;\;\;\;\; \left. - \sum_{i=1}^{n_\gamma} w_{\gamma_i}(M^2,m_j^2)
\right]
\times \exp \left( w_{\gamma_{\rm expt}} (s,m_i,\mu,\mu_{\rm expt})
\right)
\end{eqnarray}
where $n_g$ denotes the number of transversely polarized gauge bosons
and $n_f$ the number of {\it external} fermions.
This expression omits all RG corrections, even at the one loop level.
The functions $W$ and $w$ correspond to the logarithmic probability to emit
a soft and/or collinear particle per line, where the capital letters denote
the probability in the high energy effective theory and the lower case letter the
corresponding one from pure QED corrections below the weak scale. The matching
condition is implemented such that for $\mu=M$ only the high energy 
solution remains.
For the contribution from scalar fields $\phi=\{\phi^\pm,\chi,H\}$ above the
scale $M$ we have
\begin{eqnarray}
W_{\phi_i}(s,M^2) &=&  \frac{\alpha}{4 \pi}  \left[  \left( T_i(T_i+1)+  \tan^2 
\theta_{\rm w}
\frac{Y^2_i}{4} \right)  \left( \log^2  \frac{s}{M^2}- 4 \log \frac{s}{M^2}
 \right) \right. \nonumber \\ && \;\;\;\;\;\;\;\;\;\;\;\;
\left. +  \frac{3}{2} \frac{m^2_t}{M^2} \log \frac{s}{m_t^2} \right] 
\end{eqnarray}
where $\alpha=g^2/4\pi$ and $\tan^2 \theta_{\rm w}= \alpha^\prime / \alpha$.
The last term is written as a logarithm containing the top quark mass $m_t$ rather than
the weak scale $M$ since these terms always contain $m_t$ as the heaviest mass in the
loop correction \cite{m3}.
For fermions we have:
\begin{eqnarray}
W_{f_i}(s,M^2) &=&  \frac{\alpha}{4 \pi} \!\! \left[ \! \left( T_i(T_i+1)+  \tan^2 \!
\theta_{\rm w}
\frac{Y^2_i}{4} \right) \!\!
\left( \log^2 \frac{s}{M^2}- 3 \log \frac{s}{M^2}
\! \right) \right. \nonumber \\ && \left.
+ \left( \frac{1+\delta_{f,{\rm R}}}{4} \frac{m^2_f}{M^2} + \delta_{f,{\rm L}}
\frac{m^2_{f^\prime}}{4 M^2} \right)
\log \frac{s}{m_t^2} \right]
\end{eqnarray}
where $f^\prime$ denotes the weak isospin partner of $f$.
For external transversely polarized gauge bosons:
\begin{eqnarray}
W_{g_i}(s,M^2) &=& \left( \frac{\alpha}{4 \pi}T_i(T_i+1)+ \frac{\alpha^\prime}{4 \pi}
\left( \frac{Y_i}{2} \right)^2 \right) \log^2 \frac{s}{M^2} \nonumber \\
&&
- \left( \delta_{i,{\rm W}} \frac{\alpha}{\pi} \beta_0 + \delta_{i,{\rm B}}
\frac{\alpha^\prime}{\pi} \beta^\prime_0 \right) \log \frac{s}{M^2}
\end{eqnarray}
with
\begin{equation}
\beta_0=\frac{11}{12}C_A - \frac{1}{3}n_{gen}-\frac{1}{24}n_{h} \;\;\;,\;\;\;
\beta^\prime_0= - \frac{5}{9}n_{gen} -\frac{1}{24}n_{h}
\end{equation}
where $n_{gen}$ denotes the number of fermion generations and $n_h$ the number of
Higgs doublets. Again we note that for external photon and Z-boson states we must include
the mixing appropriately as discussed in Ref. \cite{m1}.
For the terms entering from contributions below the weak scale we have for fermions:
\begin{equation}
w_{f_i}(s,\mu^2) = \left\{ \begin{array}{lc} \frac{e_i^2}{(4 \pi)^2} \left( \log^2 \frac{s}{\mu^2}
- 3 \log \frac{s}{\mu^2} \right) & , \;\;\; m_i \ll \mu \\
\frac{e_i^2}{(4 \pi)^2} \left[ \left( \log \frac{s}{m_i^2}-1 \right) 2 \log \frac{m_i^2}{\mu^2}
\right. \\
\left.\;\;\;\;\;\;\;\;\;\;+ \log^2 \frac{s}{m_i^2} - 3 \log \frac{s}{m_i^2} \right] & , \;\;\; \mu
\ll m_i\end{array} \right.
\end{equation}
Analogously, for external W-bosons and photons we find:
\begin{equation}
w_{{\rm w}_i}(s,\mu^2) =
\frac{e_i^2}{(4 \pi)^2} \left[ \left( \log \frac{s}{M^2}-1 \right)
2 \log \frac{M^2}{\mu^2}
+ \log^2 \frac{s}{M^2} \right]
\end{equation}
\begin{equation}
w_{\gamma_i}(M^2,\mu^2) = \left\{ \begin{array}{lc}
\frac{1}{3} \sum_{j=1}^{n_f} \frac{e_j^2}{4 \pi^2} N^j_C
\log \frac{M^2}{\mu^2} & , \;\;\; m_j \ll \mu \\
\frac{1}{3} \sum_{j=1}^{n_f} \frac{e_j^2}{4 \pi^2} N^j_C \log \frac{M^2}{m_j^2}
& , \;\;\; \mu \ll m_j\end{array} \right.
\end{equation}
for the virtual corrections. 
For real photon emission we have in the soft
photon approximation:
\begin{eqnarray}
w_{\gamma_{\rm expt}}(s,m_i,\mu,\mu_{\rm expt})
\!\!&=&\!\!\! \left\{ \begin{array}{lc}
\sum_{i=1}^n \frac{e_i^2}{(4 \pi)^2} \left[
- \log^2 \frac{s}{\mu^2_{\rm expt}}
+ \log^2 \frac{s}{\mu^2}- 3 \log \frac{s}{\mu^2} \right]
& , m_i \ll \mu \\
\sum_{i=1}^n \frac{e_i^2}{(4 \pi)^2} \left[ \left( \log
\frac{s}{m_i^2} -  1 \right)
2 \log \frac{m_i^2}{\mu^2} + \log^2 \frac{s}{m_i^2}
\right. \\ \left. - 2 \log \frac{s}{\mu^2_{\rm expt}} \left( \log \frac{s}{m_i^2} -
1 \right) \right]
& ,
\mu \ll m_i \end{array} \right.  \\ &&
\nonumber
\end{eqnarray}
where $n$ is the number of external lines
and the upper case applies only to fermions since for $W^\pm$
we have $\mu < M$. Note that in all contributions from the regime $\mu<M$ we have
kept mass terms inside the logarithms. This approach is valid in the entire Standard
Model up to terms of order ${\cal O} \left( \log \frac{m_t}{M} \right)$.
The overall $\mu$-dependence in the semi-inclusive cross section cancels and we only have
a dependence on the parameter $\mu_{\rm expt}$ related to the experimental energy resolution.
All universal electroweak Sudakov corrections at DL and SL level exponentiate.

\subsection{Renormalization group improvement}

The way to implement the SL-RG corrections is clear from the discussion in section
\ref{sec:qcd}. At high energies, 
the DL phase space is essentially described by
an unbroken $SU(2)\times U(1)$ theory in which we can calculate the high energy contributions.
In this regime, all particle masses can be neglected and we have to consider the following
virtual electroweak DL phase space integral with running couplings in each gauge group:
\begin{eqnarray}
{\widetilde W}^{RG}_{i_V} \left(s,\mu^2 \right) &=&  \frac{1}{2\pi} 
\int^s_{\mu^2} \frac{d {\mbox{\boldmath $k$}_{\perp }^{2}}}{
{\mbox{\boldmath $k$}_{\perp }^{2}}} \int^1_{{\mbox{\boldmath $k$}_{\perp }^{2}}/s}
\frac{d v}{v}  \left\{
 \frac{T_i(T_i+1)\alpha(\mu^2)}{1+c \; \log \frac{{\mbox{\boldmath $k$}_{\perp }^{2}}}{\mu^2}} 
 + \frac{(Y^2_i/4) \alpha^\prime(\mu^2)}{1+c^\prime \; \log 
 \frac{{\mbox{\boldmath $k$}_{\perp }^{2}}}{\mu^2}} \right\} \nonumber \\
&=& \frac{\alpha(\mu^2) T_i(T_i+1)}{2 \pi } \left\{ \frac{1}{c} \log \frac{s}{\mu^2}
\left( \log \frac{\alpha(\mu^2)}{\alpha
(s)} - 1 \right) + \frac{1}{c^2}
\log \frac{\alpha(\mu^2)}{\alpha(s)} \right\} \nonumber \\
&& +\frac{\alpha^\prime(\mu^2) Y^2_i}{8 \pi } \left\{ \frac{1}{c^\prime} \log \frac{s}{\mu^2}
\left( \log \frac{\alpha^\prime(\mu^2)}{\alpha^\prime
(s)} - 1 \right) + \frac{1}{{c^\prime}^2}
\log \frac{\alpha^\prime(\mu^2)}{\alpha^\prime(s)} \right\} \label{eq:vewrg}
 \end{eqnarray}
where $\alpha(\mu^2)=g^2(\mu^2)/4 \pi$, $\alpha^\prime(\mu^2)={g^\prime}^2(\mu^2)/4 \pi$,
$c=\alpha(\mu^2)\beta_0 / \pi$ 
and analogously, $c^\prime=\alpha^\prime(\mu^2)\beta^\prime_0 / \pi$. 
In each case, the correct non-Abelian or Abelian limit is reproduced by letting the corresponding
couplings of the other gauge group approach zero. In this way it is easy to see that the
argument of the running couplings can only be what appears in Eq. (\ref{eq:vewrg}).

The form of Eq. (\ref{eq:vewrg}) is valid for fermions, transversely and longitudinally polarized
external lines but (omitted) subleading terms as well as the quantum numbers of the weak isospin $T_i$
and the weak hypercharge $Y_i$ differ. 
In order to implement the missing soft photon contribution, we choose the
analogous form of solution in Eq. (\ref{eq:sp}) and have to implement it in such a way that
for $\mu=M$ Eq. (\ref{eq:vewrg}) is obtained.
The full result for the respective semi-inclusive cross sections is then given 
by:
\begin{eqnarray}
&& d\sigma^{\rm RG} (p_{1}, \ldots, p_{n},g,g^\prime,\mu_{exp}) = d\sigma_{\rm Born} (p_{1},
\ldots ,p_{n},g(s),g^\prime (s))
\nonumber \\ && \times \exp \left\{ - \sum^{n_g}_{i=1} W^{\rm RG}_{g_i} (s,M^2)
- \sum^{n_f}_{i=1} W^{\rm RG}_{f_i} (s,M^2) - \sum^{n_\phi}_{i=1} W^{\rm RG}_{\phi_i} (s,M^2)
\right\} \nonumber \\
&&\times \exp \left[ - \sum_{i=1}^{n_f} \left( w^{\rm RG}_{f_i}(s,\mu^2)
- w^{\rm RG}_{f_i}(s,M^2) \right)
- \sum_{i=1}^{n_{\rm w}} \left( w^{\rm RG}_{{\rm w}_i}(s,\mu^2)
- w^{\rm RG}_{{\rm w}_i}(s,M^2) \right) \right. \nonumber \\
&& \;\;\;\;\;\;\;\;\;\;\; \left. - \sum_{i=1}^{n_\gamma} w_{\gamma_i}(M^2,m_j^2)
\right]
\times \exp \left( w^{\rm RG}_{\gamma_{\rm expt}} (s,m_i,\mu,\mu_{\rm expt})
\right) \label{eq:si}
\end{eqnarray}
where $n_f$ denotes here again the number of {\it external} fermions.
The argument of the gauge couplings in the Born cross section indicate the
one loop renormalization of the couplings which is not included in the exponential expressions
but which at one loop is genuinely subleading:
\begin{eqnarray}
\alpha (s) &=& \alpha (M^2) \left( 1 - \beta_0 \frac{\alpha (M^2)}{\pi} \log \frac{s}{M^2} \right) \\
\alpha^\prime (s) &=& \alpha^\prime (M^2) \left( 1 - \beta^\prime_0 \frac{\alpha^\prime
(M^2)}{\pi}\log \frac{s}{M^2} \right)
\end{eqnarray}
where $\alpha (M^2)= e^2(M^2) / 4 \pi s_{\rm w}^2$ and $\alpha^\prime (M^2)= 
e^2(M^2) / 4 
\pi c_{\rm w}^2$ with
\begin{equation}
e^2(M^2)=e^2 \left(1+ \frac{1}{3} \frac{e^2}{4 \pi^2} \sum_{j=1}^{n_f
} Q_j^2 N^j_C \log
\frac{M^2}{m_j^2} \right)
\end{equation}
and $e^2/ 4 \pi = 1/137$.
If there are non-suppressed mass ratios in the Born term, also these terms need to be renormalized
at one loop (see Ref. \cite{dp}). 
Higher order mass renormalization terms would then be sub-subleading.
The function $W^{\rm RG}_{\phi_i} (s,M^2)$ is given by
\begin{eqnarray}
 W^{\rm RG}_{\phi_i}(s,M^2)\!\!\!\!&=&\!\!\!\! \frac{\alpha(M^2) 
 T_i(T_i+1)}{2 \pi } \left\{ \frac{1}{c} \log \frac{s}{M^2}
\left( \log \frac{\alpha(M^2)}{\alpha
(s)} - 1 \right) + \frac{1}{c^2}
\log \frac{\alpha(M^2)}{\alpha(s)} \right\} \nonumber \\
\!\!\!\!&&\!\!\!\!\!\!\!\!\!\!\!\!\!\!\!\!\!\!\!\! +\frac{\alpha^\prime(M^2) Y^2_i}{8 \pi } \left\{ \frac{1}{c^\prime} \log \frac{s}{M^2}
\left( \log \frac{\alpha^\prime(M^2)}{\alpha^\prime
(s)} - 1 \right) + \frac{1}{{c^\prime}^2}
\log \frac{\alpha^\prime(M^2)}{\alpha^\prime(s)} \right\} \nonumber \\
\!\!\!\!&&\!\!\!\!\!\!\!\!\!\!\!\!\!\!\!\!\!\!\!\! - \left[ \left( \frac{ \alpha(M^2)}{4 \pi}  T_i(T_i+1)+  
\frac{ \alpha^\prime(M^2)}{4 \pi} \frac{Y^2_i}{4} \right)  4 \log \frac{s}{M^2}
- \frac{3}{2} \frac{ \alpha(M^2)}{4 \pi} \frac{m^2_t}{M^2} \log \frac{s}{m_t^2} \right] \label{eq:WRGphi}
 \end{eqnarray}
where we again have $m_t$ in the argument of the Yukawa enhanced correction \cite{m3}.
Analogously for fermions we have:
\begin{eqnarray}
W^{\rm RG}_{f_i}(s,M^2)&=& \frac{\alpha(M^2) 
 T_i(T_i+1)}{2 \pi } \left\{ \frac{1}{c} \log \frac{s}{M^2}
\left( \log \frac{\alpha(M^2)}{\alpha
(s)} - 1 \right) + \frac{1}{c^2}
\log \frac{\alpha(M^2)}{\alpha(s)} \right\} \nonumber \\
&& +\frac{\alpha^\prime(M^2) Y^2_i}{8 \pi } \left\{ \frac{1}{c^\prime} \log \frac{s}{M^2}
\left( \log \frac{\alpha^\prime(M^2)}{\alpha^\prime
(s)} - 1 \right) + \frac{1}{{c^\prime}^2}
\log \frac{\alpha^\prime(M^2)}{\alpha^\prime(s)} \right\} \nonumber \\
&& - \left[ \left( \frac{ \alpha(M^2)}{4 \pi} T_i(T_i+1)+  
\frac{ \alpha^\prime(M^2)}{4 \pi}\frac{Y^2_i}{4} \right)  3 \log \frac{s}{M^2}
\right. \nonumber \\ && \left.
- \frac{ \alpha(M^2)}{4 \pi} \left( \frac{1+\delta_{f,{\rm R}}}{4} \frac{m^2_f}{M^2} + \delta_{f,{\rm L}}
\frac{m^2_{f^\prime}}{4 M^2} \right)
\log \frac{s}{m_t^2} \right] \label{eq:WRGf}
\end{eqnarray}
The last term contributes only for left handed bottom and for top quarks as mentioned above and
$f^\prime$ denotes the corresponding isospin partner for left handed fermions.
\begin{eqnarray}
&&  W^{\rm RG}_{g_i}(s,M^2)= \frac{\alpha(M^2) 
 T_i(T_i+1)}{2 \pi } \left\{ \frac{1}{c} \log \frac{s}{M^2}
\left( \log \frac{\alpha(M^2)}{\alpha
(s)} - 1 \right) + \frac{1}{c^2}
\log \frac{\alpha(M^2)}{\alpha(s)} \right\} \nonumber \\
&& +\frac{\alpha^\prime(M^2) Y^2_i}{8 \pi } \left\{ \frac{1}{c^\prime} \log \frac{s}{M^2}
\left( \log \frac{\alpha^\prime(M^2)}{\alpha^\prime
(s)} - 1 \right) + \frac{1}{{c^\prime}^2}
\log \frac{\alpha^\prime(M^2)}{\alpha^\prime(s)} \right\} \nonumber \\
&&- \left( \delta_{i,{\rm W}} \frac{\alpha(M^2)}{\pi} \beta_0 + \delta_{i,{\rm B}}
\frac{\alpha^\prime(M^2)}{\pi} \beta^\prime_0 \right) \log \frac{s}{M^2} \label{eq:WRGg}
\end{eqnarray}
Again we note that for external photon and Z-boson states we must include
the mixing appropriately as discussed in Ref. \cite{m1}.
For the terms entering from contributions below the weak scale we have for fermions:
\begin{equation}
w^{\rm RG}_{f_i}(s,\mu^2)\!\! = \!\!\left\{ \begin{array}{lc} \!\!\!
\frac{e^2_i}{8 \pi^2 } \left\{ \frac{1}{c} \log \frac{s}{\mu^2}
\left( \log \frac{e^2(\mu^2)}{e^2
(s)} - 1 \right) + \frac{1}{c^2}
\log \frac{e^2(\mu^2)}{e^2(s)} 
- \frac{3}{2} \log \frac{s}{\mu^2} \right\} \!\!& , \; m_i \ll \mu \\ \!\!\!
= \frac{e_i^2}{8 \pi^2 } \left\{ \frac{1}{c} \log \frac{s}{m^2}
\left( \log \frac{e^2(\mu^2)}{e^2
(s)} - 1 \right) -\frac{3}{2} \log \frac{s}{m^2} - \log \frac{ m^2}{\mu^2} \right. \\ 
\left. + \frac{1}{c^2}
\log \frac{e^2(m^2)}{e^2(s)} \left( 1- \frac{1}{3}\frac{e^2}{4 \pi^2}
\sum_{j=1}^{n_f } Q_j^2 N^j_C \log \frac{m^2}{m_j^2}
\right) \right\}
\!\!& , \; \mu
\ll m_i\end{array} \right.
\end{equation}
where $c=-\frac{1}{3}\frac{e^2}{4 \pi^2} \sum^{n_f}_{j=1}Q^2_j N_C^j$.
Analogously, for external W-bosons and photons we find:
\begin{eqnarray}
w^{\rm RG}_{{\rm w}_i}(s,\mu^2) &=& 
\frac{e_i^2}{8 \pi^2 } \left\{ \frac{1}{c} \log \frac{s}{M^2}
\left( \log \frac{e^2(\mu^2)}{e^2
(s)} - 1 \right) - \log \frac{ M^2}{\mu^2} \right. \\ 
&& \left. + \frac{1}{c^2}
\log \frac{e^2(M^2)}{e^2(s)} \left( 1- \frac{1}{3}\frac{e^2}{4 \pi^2}
\sum_{j=1}^{n_f } Q_j^2 N^j_C \log \frac{M^2}{m_j^2}
\right) \right\}
\end{eqnarray}
\begin{equation}
w_{\gamma_i}(M^2,\mu^2) = \left\{ \begin{array}{lc}
\frac{1}{3} \sum_{j=1}^{n_f} \frac{e_j^2}{4 \pi^2} N^j_C
\log \frac{M^2}{\mu^2} & , \;\;\; m_j \ll \mu \\
\frac{1}{3} \sum_{j=1}^{n_f} \frac{e_j^2}{4 \pi^2} N^j_C \log \frac{M^2}{m_j^2}
& , \;\;\; \mu \ll m_j\end{array} \right.
\end{equation}
Note that the function $w_{\gamma_i}(M^2,\mu^2)$ does not receive
any RG corrections to the order we are working since it contains only SL terms.
For the virtual corrections and for real photon emission we have in the soft
photon approximation:
\begin{eqnarray}
w^{\rm RG}_{\gamma_{\rm expt}}(s,m_i,\mu,\mu_{\rm expt})
\!\!&=&\!\!\! \left\{ \begin{array}{lc}
\sum_{i=1}^n \frac{e_i^2}{8 \pi^2} 
 \left\{ \frac{1}{c} \log \frac{s}{\mu^2} \left( \log \frac{
e^2(\mu^2)}{e^2 (\mu^2_{\rm expt})} - 1 \! \right) - \frac{1}{c} \log \frac{\mu^2_{\rm expt}}{
s} \right. \\ \left.
+ \frac{1}{c^2} \log \frac{e^2 (\mu^2)}{e^2 (\mu^2_{\rm expt})} 
 -\frac{3}{2} \log \frac{s}{\mu^2}
\right\} 
\;\;\;\;\;\;\;\;\;\;\;\;\;\;\;\;\;\;\;\;\;\;\;\;\;\;\;\;\;\;\;\;\; , m_i \ll \mu \\
\sum_{i=1}^n \frac{e_i^2}{8 \pi^2} 
 \left\{ \frac{1}{c} \log \frac{s}{m^2}
 \left( \log \frac{ e^2 (\mu^2)}{e^2 (\mu_{\rm expt}^2m^2/s)} - 1  \right) \right. 
\\ \!\!\!  + \frac{1}{c^2} \log \frac{e^2(\mu^2_{\rm expt}m^2/s)}{e^2(\mu_{\rm expt}^2)} 
\left( 1-
 \frac{1}{3} \frac{e^2}{4 \pi^2} \sum^{n_f}_{j=1}
 Q^2_j N_C^j
\log \frac{
\mu_{\rm expt}^2}{m_j^2} \right)
\\ \left.
- \log \frac{m^2}{\mu^2} + \log \frac{s}{\mu_{\rm expt}^2}
\right\} 
\;\;\;\;\;\;\;\;\;\;\;\;\;\;\;\;\;\;\;\;\;\;\;\;\;\;\;\;\;\;\;\;\;\;\;\;\;\;\;\;\;\;\;
,\mu \ll m_i \end{array} \right. \nonumber \\ &&
\end{eqnarray}
where $n$ is the number of external lines and $n_f$ fermions propagating in
the loops folded with the DL integrals.
The upper case applies only to fermions since for $W^\pm$
we have $\mu < M$. Note that in all contributions from the regime $\mu<M$ we have
kept mass terms inside the logarithms. 
For the running above the weak scale $M$ we use only the massless $\beta_0$, $\beta_0^\prime$
terms with $n_{gen}=3$. This approach is valid in the entire Standard
Model up to terms of order ${\cal O} \left( \log \frac{m_t}{M} \right)$.

\section{Discussion} \label{sec:dis} 

\begin{figure}
\centering
\epsfig{file=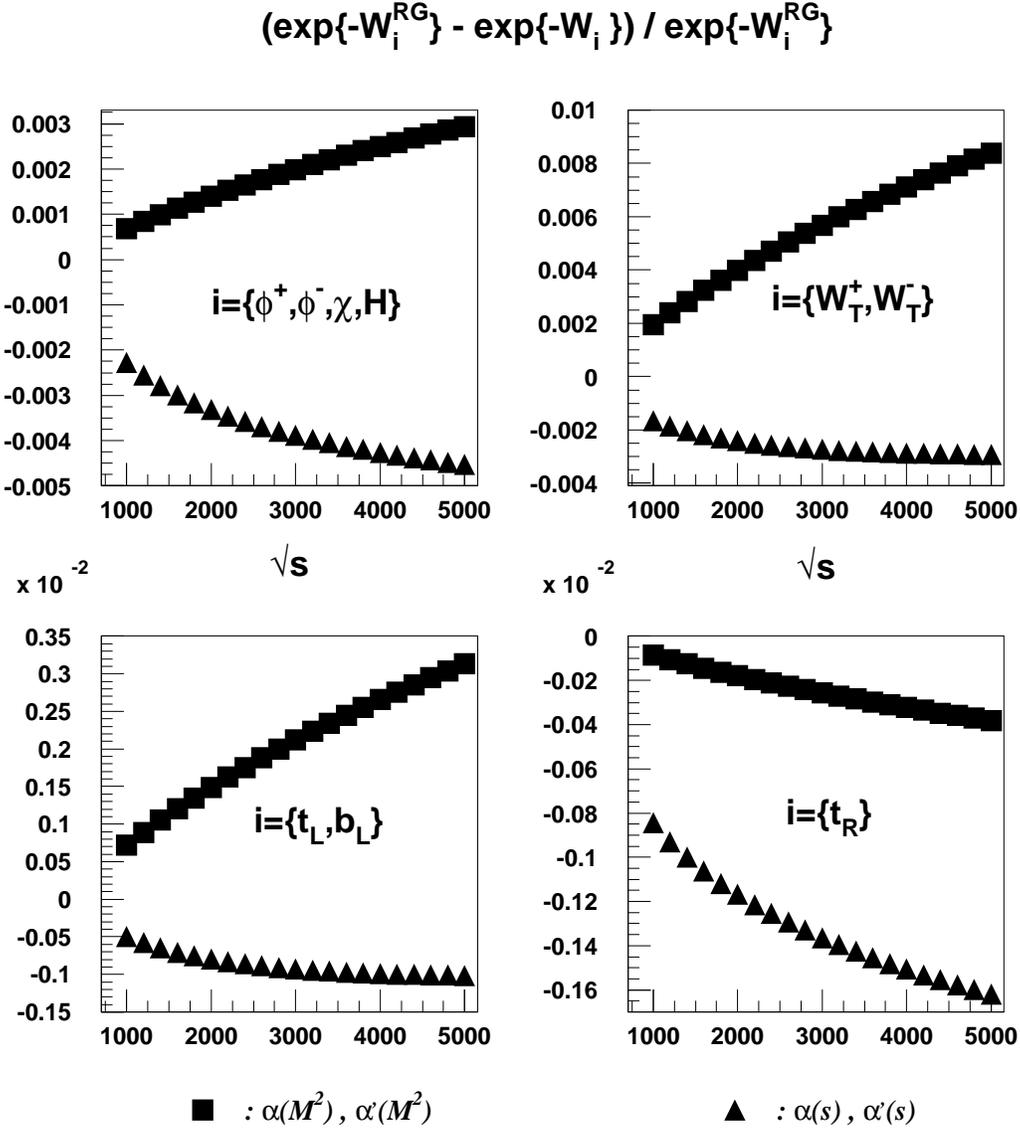,width=16cm}
\caption{This figure compares the renormalization group improved probabilities $W^{\rm RG}_i$
with the conventional Sudakov exponentials $W_i$ for various external particle lines. The
comparison is made with the indicated scale choices for the functions $W_i$ and takes into
account only the RG corrections from the scale $M$ to $\sqrt{s}$. 
Taking the difference between the two curves is a measure of the uncertainty removed 
in this work.
The variations in 
the scale of the coupling in the $W_i$ functions is largest in the scalar (Goldstone and
Higgs boson) sector
and for transverse $W^\pm$ where the effect is
about $0.6\%$ at 2 TeV per line on the level of the cross section. In general, the RG
improved form factors differ by fractions of one percent per line and need to be taken into
account at future colliders if the experimental accuracy is in the percentile regime.}
\label{fig:wrel}
\end{figure}

In this section we discuss briefly the size of the SL-RG corrections obtained in section
\ref{sec:sm}. For this purpose we will only compare the terms which are
new in the present analysis, i.e. the running from the weak scale $M$ to $\sqrt{s}$. We
are thus interested in effects starting at the two loop level and want to compare the relative
size of the RG-improved form factors to the pure Sudakov terms. It is therefore of interest
to compare the ratios $\left(e^{\{-W^{\rm RG}_i\}}-e^{\{-W_i\}} \right)/ e^{\{-W^{\rm RG}_i\}}$
for the various particle labels $i$. Since the physical scales in the problem are given
by $M$ and $\sqrt{s}$, the lower and upper limits of the couplings are given accordingly
by these scales for the functions $W_i$. Fig. \ref{fig:wrel} compares the respective 
ratios for various SM particles. 
For definiteness we take $M=80$ GeV, $m_t=175$ GeV, $s_{\rm w}^2=0.23$, 
$\alpha (M^2)=1/128/s_{\rm w}^2$, $\alpha^\prime (M^2)=1/128/c_{\rm w}^2$, $\beta_0
=19/24$ and $\beta_0^\prime=-41/24$.
The difference between the curves using $M^2$ and those
using $s$ as the scales in the conventional Sudakov form factors is a measure of the
inherent scale uncertainty which is removed by the RG-improved Sudakov form factors
$W^{\rm RG}_i$. The largest effect is obtained in the gauge boson sector.
For external $\{ \phi^+,
\phi^-, \chi, H \}$ particles we have at 1 TeV a difference between the curves of
about $0.3\%$ per line on the level of the cross section, growing to $0.5\%$ at 2 TeV.
The situation is very similar for transversely polarized $W^+, W^-$ particles where
it reaches about 0.35\% at 1 TeV and 0.6\% at 2 TeV per line on the cross section
level. 
For left handed quarks of the third generation the size of the corrections is about
0.125\% at 1 TeV per line on the level of the cross section and 0.25\% at 2 TeV. These corrections
are thus considerably smaller and only needed if precisions below the one percent level are
necessary from the theory side. For right handed top quarks the effect is even smaller 
since only the running of $\alpha^\prime$ enters and it is
thus negligible for most applications. The form of the two curves in case of right handed
tops differs markedly from the other three cases because at the energies displayed, the 
dominant effect is actually due to subleading Yukawa enhanced corrections ($\sim
\alpha$) since the DL
terms are proportional to $\alpha^\prime$ and since the ratio $m_t^2/M^2$ is of the size of
an additional logarithm for these values of $\sqrt{s}$.

In general it can be seen that-where the DL terms dominate-the renormalization group
improved results are indeed in-between the upper and lower bounds given by the respective
scale choices in the conventional Sudakov form factors. Indeed also for right handed top
quarks this pattern is observed if only DL corrections are taken into account.

It should be emphasized again that also the QED-RG corrections can be sizable since large mass
ratios with light particles occur. These should of course also be implemented in a full
SM prediction at TeV energies.

\section{Conclusions} \label{sec:con}

In this paper we have obtained the complete subleading
electroweak renormalization group corrections to all orders in high energy processes
in the framework of the infrared evolution equation (IREE) method. These
are terms originating in loops which at a lower loop order lead to DL corrections and are of
the type $\alpha^n \beta_0 \log^{2n-1} \frac{s}{M^2}$. We have derived the corrections for
massless as well as massive gauge theories and used appropriate matching conditions to 
obtain the full SM contributions.
These corrections start at the two loop level and are universal, i.e. properties of external
lines and thus process independent.
They represent the last missing universal contribution needed for a full SL-analysis at
the two loop level. The size of the effect at TeV energies is in general a fraction of
one percent and is largest in the scalar and gauge boson sector, where at 2 TeV the uncertainty
in the conventional Sudakov form factor is about 0.6\% per line at the level of the cross section.
These effects cannot be neglected at TeV linear colliders for precision measurements in the
percentile regime. 

The last outstanding type of SL-correction at the two loop level
is given by the non-universal, process dependent
angular terms of the type $\alpha^n \log^{2n-1} \frac{s}{M^2} \log \frac{u}{t}$. 
These terms also don't factorize
with respect to the Born cross amplitude and the high precision objectives of future linear
colliders will make at least a two loop analysis of these corrections mandatory.

\section*{Acknowledgments}

I would like to thank A.~Denner for carefully reading the manuscript.

\end{document}